\documentstyle[epsf,12pt]{article}
\newcommand{\al}{\alpha'}
\newcommand{\de}{\partial}
\newcommand{\be}{\begin{equation}}
\newcommand{\ba}{\begin{eqnarray}}
\newcommand{\ea}{\end{eqnarray}}
\newcommand{\ee}{\end{equation}}
\newcommand{\db}{\bar{\partial}}

\newcommand{\ca}{\mathcal}

\newcommand{\f}{\frac}
\newcommand{\s}{\sqrt}
\newcommand{\vp}{\varphi}

\newcommand{\mb}{\mathbf}

\newcommand{\no}{\nonumber \\}

\newcommand{\ddbt}{\mbox{D}2-\overline{\mbox{D}2}}

\begin{document}
\begin{titlepage}
\thispagestyle{empty}
\begin{flushright}
hep-th/0112199 \\
UT-980 \\
December, 2001 \\
\end{flushright}

\bigskip
\bigskip

\begin{center}
\noindent{\Large \textbf{Flux Stabilization of D-branes\\ 
\bigskip
in NSNS Melvin Background}}\\
\bigskip
\bigskip
\bigskip
\noindent{
          Tadashi Takayanagi\footnote{
                 E-mail: takayana@hep-th.phys.s.u-tokyo.ac.jp}
                 and Tadaoki Uesugi\footnote{
                 E-mail: uesugi@hep-th.phys.s.u-tokyo.ac.jp} }\\

\bigskip
\it{Department of Physics, Faculty of Science \\ University of Tokyo \\
\medskip
Tokyo 113-0033, Japan}

\end{center}
\bigskip
\begin{abstract}
In this paper we reexamine the D-brane spectrum in the Melvin background
with non-constant NSNS B-field from the viewpoint of its world-volume and 
string world-sheet theory. 
We find that the stable D2-D0 bound state exists even 
though it does not wrap any nontrivial cycles. We show that this
system is stabilized by the presence of the NSNS B-field and the 
magnetic flux $F$. Moreover from the non-abelian world-volume theory of 
D0-branes the bound state is regarded as a system of D0-branes expanding 
into a fuzzy torus. 
\end{abstract}
\end{titlepage}

\newpage

\section{Introduction}
\setcounter{equation}{0}
\hspace{5mm}

Recently many aspects of Melvin backgrounds \cite{Me} 
in string theory have been 
studied intensively.
They give us interesting models of string theory in flux backgrounds. 
In the case
of RR fluxes they are often called fluxbranes \cite{CG,Sa,GS,RT,CoHeCo,Em,Ur,MoTr} and 
one can realize them as classical solutions in supergravity theory. 
On the other hand,
if we consider NSNS fluxes, then one can have exactly solvable string sigma
models \cite{BM3,BM,SM,TU1,RuTs}. 

The latter is regarded as a ${\bf S}^1$ fibration over ${\bf R}^2$. 
This non-trivial fibration is due to 
two Kaluza-Klein (K.K.)
 gauge fields $A_{\vp}$
and $B_{\vp}$ , 
which originate from K.K. reduction of metric 
$G_{\vp y}$ and B-field $B_{\vp y}$, respectively. 
The metric of this background $\mbox{M}_3$ is given by
\ba
\label{metric}
ds^2&=&d\rho ^2+\f{\rho ^2}{(1+\beta^2\rho^2)(1+q^2\rho^2)}d\vp^2+
\f{1+q^2\rho^2}{1+\beta^2\rho^2}(dy+A_{\vp}d\vp)^2, \no 
A_{\vp}&=&\f{q\rho^2}{1+q^2\rho^2},\ \ \ 
B_{\vp y}\equiv B_{\vp}=-\f{\beta\rho^2}{1+\beta^2\rho^2},\ \ \ 
e^{2(\phi-\phi_0)}
=\f{1}{1+\beta^2\rho^2},  \label{KK}
\ea
where $q, \beta$ are the (magnetic) parameters which are proportional to the 
strength of two gauge fields and $\phi_0$ is the constant value of the 
dilaton $\phi$ at $\rho =0$. As a string model we can consider the
ten dimensional background ${\bf R^{1,6}}\times \mbox{M}_3$. 

Since
the presence of fluxes generically breaks all supersymmetries, 
the theory becomes
unstable and often includes closed string tachyons (for the recent 
analysis of
tachyon condensation in NSNS Melvin backgrounds see 
\cite{Su2,TU1,RuTs,MiYi,Minwalla,Su3}).

Another interesting phenomenon caused by the NSNS fluxes 
which we will discuss is 
the existence of expanded D2-branes (see Fig.1) 
whose world-volume has the form of 
a torus. This is accompanied with a quantized magnetic flux $F$ on it and
thus is more properly said as a D2-D0 bound state.
This kind of D2-branes has already been expected by the 
heuristic arguments on T-duality transformations in the paper \cite{TU2}.
However, we have been left with the important 
question why such a topologically trivial
torus D2-brane is stable. In this short letter we answer this by
investigating the D-brane from the viewpoint of the D-brane world-volume
theory and the would-sheet theory. We conclude that 
such a D2-brane is stabilized by the combined effect of both the 
NSNS flux $B$
and the quantized magnetic flux $F$ on it. 

\section{Flux Stabilization of D-branes in NSNS Melvin Background}
\setcounter{equation}{0}
\hspace{5mm}

Let us consider what kind of D-branes can exist in the Melvin background
(\ref{KK}). In the most part of this paper we assume that 
the values of the magnetic parameters are rational such that
\ba
\f{\beta\al}{R}=\f{k}{N},\ \ \ qR=\f{l}{M}, \label{ra}
\ea
where $(k,N)$ and $(l,M)$ are pairs of coprime integers.
In particular we are interested in such D-branes that are localized 
in the $\rho$ direction. Thus we assume that D-branes obey the Dirichlet 
boundary condition along $\rho$. The other D-branes can also be investigated
in the same way as the arguments below and we summarize the results in 
Table 1. 

First let us discuss a D0-brane. If we
put it in the Melvin background, then we can see that it can exist only at 
the origin $\rho=0$ because of the non-trivial $\rho$ dependence 
of the dilaton $\phi$ in (\ref{KK}). This is easily understood if we remember the value of
D0-brane mass $\mbox{M}_{D0}=e^{-\phi}(\al)^{-\f12}$, which takes its minimum
value at $\rho=0$.

Next we consider a D2-brane whose world-volume is the torus
$0\le \vp <2\pi,\ 0\le y< 2\pi R,\  \rho=\mbox{constant}$.
However, it is easy to see that the mass of it
is proportional to $\rho$ (set $F$ to zero in (\ref{mass}))
as in the flat space. Thus it should squash and cannot exist. 

In this way we have observed that any D2-branes and D0-branes cannot exist at
$\rho\neq 0$. Then what happens if we consider D2-D0 bound states? 
We start with a D2-D0 bound state which is made of $p$ D2-branes and 
$q$ D0-branes ($p$ and $q$ are coprime) and assume that 
its world-volume is the same torus. 
The mass\footnote{Here we ignore the quantum corrections. On the other
hand we will not have such a problem for the supersymmetric example discussed
at the next section.} of this object is given by
\ba
\label{mass}
\mbox{M}_{p,q}&=&\f{e^{-\phi}}{4\pi^2(\al)^{\f32}}\int dy\ d\vp
\mbox{Tr}\ \s{\det(G+B+F)}\no
&=&\f{e^{-\phi_0}pR}{(\al)^{\f32}}\s{(F\beta-1)^2\rho^2+F^2},
\ea
where $F$ is the constant flux which generates $q$ D0-branes and is 
quantized as usual\footnote{Note that
this value $F$ is determined by the quantization law
$\frac{1}{4\pi^2\al}\int\mbox{Tr}\ F=q\in \mb{Z}$.}
\ba
F=\al R^{-1}\ \f{q}{p}~~. \label{fl}
\ea
In order for this D-brane to exist at $\rho\neq 0$, 
the $\rho$ dependence of the energy
should disappear and we have the constraint $F=1/\beta$.
Since we assume the rational cases (\ref{ra}),
this can be satisfied for $p=k,\ q=N$. Furthermore the mass of the object
for this particular value of flux is given by
\ba
\mbox{M}_{k,N}&=& N\ \mbox{T}_{0}\ \ \ (\mbox{T}_{0}=
e^{-\phi_0}(\al)^{-\f12}), \label{mass1}
\ea
where $\mbox{T}_{0}$ is the mass of a D0-brane at $\rho=0$.
This result (\ref{mass1}) tells us an interesting fact that
the D2-brane part of the mass $\mbox{M}_{k,N}$ is effectively zero (so
called tensionless brane). 
This is the reason why such a expanded D-brane is allowed which does not 
wrap any nontrivial cycles.
Moreover from the RR-coupling and eq.(\ref{fl}) 
we can obtain the correct RR-charges of $k$ D2-branes and $N$ 
D0-branes\footnote{In RR-coupling there is a term which includes
B-field such as $\int C_{D1}\wedge B$, while this term will be canceled by
the bulk term in the supergravity action\cite{Ta}.}. This mechanism may
be regarded as a non-compact CFT version
of the stabilization of spherical D2-branes in SU(2)
WZW-model (NS5-brane background)
\cite{douglas,paw,ARS,Ha,HNS,Hyakutake}.

\begin{table}[tbp]
\begin{center}
\begin {tabular}{|c|c||c|c|}
\hline
  Free field & $(Y',\rho,\vp'')$ & Melvin $(\rho,\vp,Y)$ & Tension \\
\hline
  D0* & DDD & D0 fixed at $\rho=0$ & $T_0$ \\
  D0  & DDD & D2-D0 bound state ($\rho,Y=$fixed)& $NT_0$ \\
  D1  & DND & D3-D1 bound state & $NT_1$ \\
  D2*  & DNN & D2 ($Y=$fixed) & $T_2$ \\ 
\hline
  D1* & NDD & D1 fixed at $\rho=0$ & $T_1$ \\
  D1  & NDD & Spiral D1 ($\rho,\ \vp+qY=$fixed) & $MT_1$ \\
  D2  & NND & Spiral D2 ($\vp+qY=$fixed) & $MT_2$ \\
  D3*  & NNN & D3 & $T_3$ \\
\hline
\end{tabular}\\
	\caption{D-brane spectrum in the Melvin background with rational 
	values of parameters $\beta\al/R=k/N,qR=l/M$. We show how the 
	D-branes defined in the free field representation $(Y',\rho,\vp'')$ 
	correspond to those in the original Melvin background. In the above 
	table the Neumann and Dirichlet boundary condition are denoted by $N$ 
	and $D$. The tension $T_p$ represents that of the standard Dp-brane.
	The D-branes marked by * are regarded as fractional D-branes and 
 even for irrational case they 
	have finite tensions, while others have 
	infinite tensions for irrational case.}
	\label{D-brane1}
\end{center}
\end{table}

It is also interesting to examine the limit $\rho\to 0$.
Since the net D2-brane charge is zero for this torus configuration, we
have only $N$ D0-branes localized at $\rho=0$. This is similar to
the decay of a $\ddbt$ system due to tachyon condensation \cite{Sen}. 
However, note 
that our process $\rho\to 0$ is an exactly marginal deformation of boundary
conformal field theory as we will see. If we say these results in the 
opposite way, $N$ D0-branes with the torus D2-brane can leave from the
origin $\rho=0$. This behavior is very similar to that of
fractional D-branes in ${\bf{Z}}_N$ orbifolds \cite{DoMo,DiDoGo}. 
Indeed as has been
already pointed out in our previous paper \cite{TU2} by heuristic arguments
of T-duality transformations, 
we can identify the D2-D0 bound state in the original coordinate system
$(\rho,\vp,Y)$ as 
the system of $N$ different fractional D-branes in the other coordinate 
system $(\rho,\vp'',Y')$ (see Fig.1). 
Let us see this correspondence more explicitly. 

\begin{figure}[htbp]
     \epsfxsize=150mm
  \centerline{\epsfbox{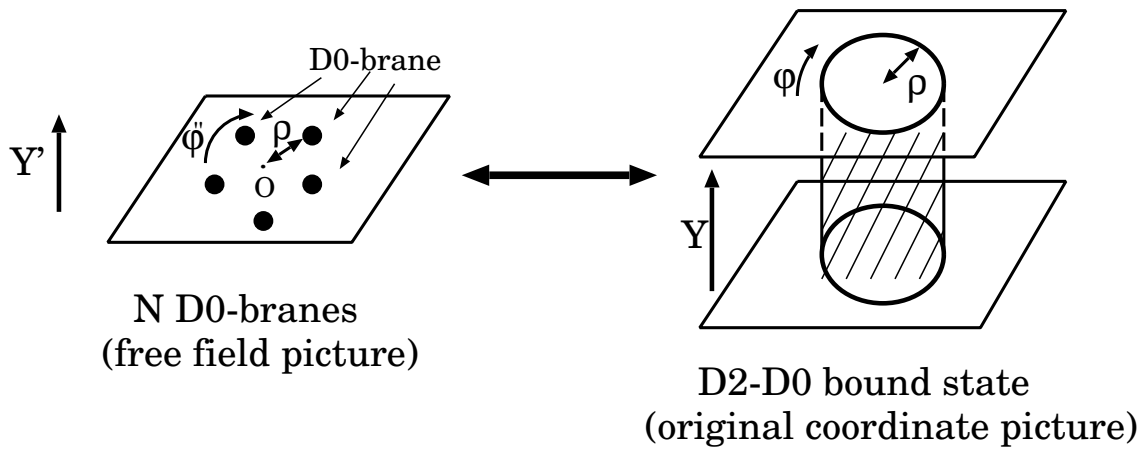}}
	\caption{The equivalence between $N$ D0-branes in the free field theory 
	and a D2-D0 bound state in the original sigma model of the Melvin 
	background.}
	\label{Fig1}
\end{figure}

As shown in \cite{BM,SM}, the sigma model of the NSNS Melvin background
(\ref{KK}) can be solvable. Indeed we can show that if we perform T-duality 
twice, then the sigma model is equivalent to that of the 
flat background with nontrivial boundary conditions 
(for more details see the review part in \cite{TU2}). 
The world-sheet fields for the trivial background, which define a free 
field theory, are denoted by
$(X'=\rho e^{i\vp''},\bar{X}'=\rho e^{-i\vp''},Y')$.
The relation between these free fields and world-sheet fields in the 
original NSNS Melvin background is given as follows
\ba
\label{transform}
(1+\beta^2\rho^2)\de \vp''&=&\de(\vp+qY)+\beta\de Y, \no
(1+\beta^2\rho^2)\db \vp''&=&\db(\vp+qY)-\beta\db Y, \no
(1+\beta^2\rho^2)\de Y'&=&\de Y-\beta\rho^2\de(\vp+qY), \no
(1+\beta^2\rho^2)\db Y'&=&\db Y+\beta\rho^2\db(\vp+qY).
\ea
The Dirichlet boundary conditions of D0-branes are
\ba
\de_2 \vp ''=0,\ \ \ \de_2 Y'=0,
\ea
at $\sigma_1=0,\pi$ (from now on we will define the boundary conditions in
the open string picture). If we rewrite the above equations from the
viewpoint of the original Melvin sigma model by using (\ref{transform}), 
they become
\ba
&&i\de_2 (\vp+qY)-\beta\de_1 Y=0,\no
&&i\de_2 Y+\beta\rho^2\de_1 (\vp+qY)=0. \label{nd}
\ea
Thus we have obtained mixed Neumann-Dirichlet boundary
conditions. By comparing this result (\ref{nd}) with the general
formula of the boundary condition
\ba
\label{gbc} 
G_{\mu\nu}\de_1 X^{\nu}+i(B_{\mu\nu}+F_{\mu\nu})\de_2 X^{\nu}=0,
\ea
where $X^{\mu}$ denotes the world-sheet field, we obtain the non-trivial 
value of the flux
\ba
F=F_{\vp y}=\f{1}{\beta}=\al R^{-1}\ \f{N}{k}.
\ea 
This value does match with the previous value (\ref{fl}) if we set
$p=k,\ q=N$. Therefore we can conclude that $N$ D0-branes at $\rho\neq 0$
in the free field picture in $(\rho,\vp'',Y')$ is 
T-dual equivalent to a bound state of $N$ D0-branes
and $k$ D2-branes wrapping around the torus $(\vp,Y)$ with $\rho\neq 0$
in the original coordinate picture.
This shows that expanding the D2-brane corresponds to moving the fractional
D0-branes and thus this is an exactly marginal deformation of boundary
conformal field theory.
It would be also interesting that the quantization of flux $F$ 
requires the rational values of $\f{\beta\al}{R}$. In the irrational
cases we will have to require
$N\to\infty$ in order to move D0-branes, 
and the bound state becomes infinitely massive.

Let us comment the world-volume theory on a D2-D0 bound state.
Because of the presence of B-flux it becomes noncommutative theory
\cite{CDS,SW}.
Following the prescription \cite{SW}, it is easy to see that the 
noncommutativity $\theta$ of noncommutative torus ${\ca{A}}_{\theta}$ 
is exactly given for any value of $\rho$ as follows
\ba
\theta=\f{\beta\al}{R}=\f{k}{N}\in {\bf Q}.
\ea
This shows that it is identified with the fuzzy torus which allows finite
dimensional representations.

Such a noncommutativity can be seen more explicitly from the analysis of 
the non-abelian Dirac-Born-Infeld (DBI) action of $N$ D0-branes. 
The non-abelian DBI action is already proposed in
\cite{Ts0,My1} where the author determines its form from the T-duality
covariance of DBI action. Especially the action of $N$ D0-branes is
written by
\ba
S_{DBI}=-T_0\int dt~ {\rm STr}\left[e^{-\phi}\sqrt{-P\{E_{00}
+E_{0i}(Q^{-1i}_{~~~j}-\delta^i_{~j})E^{jk}E_{k0}\}\det(Q^i_{~j})}\right],
\ea
where we defined $E_{\mu\nu}$ and $Q^i_{~j}$ as 
$E_{\mu\nu}=G_{\mu\nu}+B_{\mu\nu}~~(\mu,\nu=0,\cdots ,9)$, 
$Q^i_{~j}=\delta^i_{~j}+i[\Phi^i,~\Phi^k]E_{kj}~~(i,j,k=1,\cdots ,9)$, and 
$\rm{STr}$ and $P$ denote the symmetrized trace and 
the pull back onto the D0-brane world-volume, respectively. 
Here we do not consider the time dependence of fields, 
thus we can set $\frac{d\Phi^i}{dt}$ to zero. Moreover since the transverse 
fluctuations for ${\bf R^{1,6}}$ directions are irrelevant in this
situation, we can also set such fields to zero. 
After all, the potential part of the action becomes the following form
\ba
V= T_0 e^{-\phi_0}{\rm STr}
\left[\sqrt{1+\rho^2\left\{i[\Phi^{\varphi},~\Phi^{Y}]+\beta\right\}^2}\right].
\ea
Here we set $\Phi^{\rho}$ to $\rho$ because we want to 
consider the expanding of D0-branes into the torus form of D2-branes with 
constant $\rho~(\neq 0)$. From the above equation we
can see that the potential is always greater than $NT_0 e^{-\phi_0}$,
which is the mass of $N$ D0-branes. To realize such the lowest limit of
the potential the condition $[\Phi^{\varphi},~\Phi^{Y}]=i\beta$ should be
needed. If we normalize coordinates by replacing $\Phi^{y}$ with 
$\frac{\Phi^{y}R}{2\pi\al}$, 
then we can get the following relation
\ba
\label{noncom}
[\Phi^{\varphi},~\Phi^{Y}]=2\pi i\theta.
\ea
This is exactly the algebra of noncommutative torus. However this
relation holds only for the infinite dimensional
representation of $\Phi^{\varphi}$ and $\Phi^{Y}$, while here we consider
the finite dimensional ($N\times N$) representation. We can approximate
by using $N$ dimensional fuzzy torus algebra generated by 
$e^{i\Phi^{\vp}}=U$ and $e^{i\Phi^{Y}}=V$ with the relation $UV=e^{-2\pi 
i\frac{k}{N}}VU$. Then the potential is not exactly equal to the
mass of $N$ D0-branes\footnote{Of course this relation exactly holds for
the infinite number of D0-branes, however in that case 
the potential value becomes infinite and this configuration may be
singular. This consideration may be related to the 
D-brane picture in the Melvin background with irrational magnetic
parameters \cite{TU2}.}. Such a difference comes from the $\frac{1}{N}$ order
correction which can be seen in other non-abelian world-volume 
analyses \cite{My1, My2, Hyakutake}. Any way the
noncommutative algebra of the torus (\ref{noncom}) is a good approximation
for large $N$ 
and we have seen the explicit noncommutativity on the world-volume 
of the D2-brane. 
     
Finally let us turn to a D1-brane in the Melvin background \cite{DuMo,TU2}.  
As shown in \cite{TU2}, this is 
equivalent to the previous D2-D0 bound state by the T-duality along $Y'$,
which interchanges $q$ and $\beta$ \cite{SM, TU1}.
 The boundary condition of a D1-brane in the free
field theory can be rewritten in terms of the fields $(\rho,\vp,Y)$ and 
the result is
\ba
\de_2 (\vp+qY)=\de_1 Y=0.
\ea
This exactly represents a D1-brane wrapping the `geodesic line for D-branes' 
$\vp+qY=\mbox{constant}$, which is defined\footnote{
Note that the usual geodesic line for the metric $(ds)^2$ is given by
$\vp+(q\pm \beta)Y=\mbox{constant}$.}
for the `D-brane metric'
$e^{-2\phi}(ds)^2$.

As pointed out in the calculation of the boundary state 
\cite{TU2}, 
for rational values of $qR=\f{l}{M}$ 
the mass of the D1-brane is finite and it winds $M$ times along $Y$ 
and $l$ times along $\vp$, while for
irrational values it becomes infinite because the D1-brane should wind
infinitely many times. These facts are all consistent with the T-dual 
equivalence to the previous D2-D0 bound state.

\section{D-branes in Higher Dimensional Melvin Background}
\setcounter{equation}{0}
\hspace{5mm}

In previous section we considered the D2-D0 bound state in the Melvin
background. However, in the presence of the flux 
this background breaks the target space supersymmetry completely, and 
the D2-D0 bound system is not a BPS state.  
 
On the other hand, as shown in several papers \cite{TU1, RuTs}, we can extend the exactly solvable NSNS Melvin background 
${\bf R}^{1,6}\times \mbox{M}_3$ 
to more higher dimensional ones ${\bf R}^{1,8-2n}\times \mbox{M}_{2n+1}$.
The most important nature of higher
dimensional backgrounds is that the partial supersymmetry can be preserved even
though these backgrounds are curved. Thus, we can consider BPS
D-branes in these backgrounds. They are stable, and
the classical analysis is reliable. 
Therefore it is interesting to consider D-branes in 
these higher dimensional backgrounds. 

Here we consider one example of supersymmetric higher dimensional
models; the product space of the trivial five dimensional Minkowski
Space ${\bf R^{1,4}}$ and the nontrivial curved space $\mbox{M}_5$ which is 
topologically equivalent to ${\bf S^1}$ fibration over 
${\bf R^2}\times{\bf R^2}$. In the following we parameterize 
${\bf S^1}$ and  ${\bf R^2}\times{\bf R^2}$ 
with $Y$ and $(\rho, \varphi), (r,\theta)$, respectively. Then, 
its world-sheet action is given as follows \cite{TU2}
\ba
\label{Hiaction}
S&=&\f{1}{\pi\al}\int d^2\sigma\Bigl[\db\rho\de\rho+\db r\de r+
\rho^2\db\check{\vp}\de\check{\vp}+r^2\db\check{\theta}\de\check{\theta}\no
&&+(1+\beta_1^2\rho^2+\beta_2^2 r^2)^{-1}(\db Y+\beta_1\rho^2\db\check{\vp}
+\beta_2 r^2 \db\check{\theta})(\de Y-\beta_1\rho^2\de\check{\vp}
-\beta_2 r^2 \de\check{\theta})\Bigr],
\ea
where we have defined $\check{\vp}=\vp+q_1Y,\ \check{\theta}=\theta+q_2Y$.
This model includes four flux parameters $q^1,~q^2,~\beta^1$ and
$\beta^2$, and as you can see from the Killing spinor analysis 
\cite{TU1,TU2}, this model 
keeps half of the maximal supersymmetry 
if $q_1=q_2,\beta_1=\beta_2$ or $q_1=-q_2,\beta_1=-\beta_2$.

Here we consider a D2-D0 bound state 
in the higher dimensional background with rational parameters 
$\beta_i=\frac{k_i}{N}~(i=1,2)$, where $N$, $k_1$ and $k_2$ are coprime 
integers. For simplicity we set $q_i~(i=1,2)$ to zero. This system becomes
a BPS state \cite{TU2} if the background keeps supersymmetry. 
The main motivation to analyze this is the same as that in the previous 
section: 
the stabilization mechanism which comes from NSNS B-field effect. 
The analysis is almost the same as that in the previous section, while 
there appears one nontrivial constraints which we will see. 

First we examine the boundary condition of this D2-D0 bound state. 
In this case we can transform the original 
coordinates $(Y,\rho,\vp,r,\theta)$ in 
(\ref{Hiaction}) into the free fields $(Y',\rho,\vp'',r,\theta'')$ 
by using T-duality two times and several field redefinitions. 
To analyze this system quantitatively we transform the following boundary 
conditions of D0-branes
\ba
\de_2\vp''=\de_2\theta''=\de_2 Y'=0,
\ea
into those which are represented by the original coordinate 
$(\vp,\theta,Y)$. The relation between these 
coordinates are obtained in the same way as (\ref{transform}). 
Then the result becomes 
\ba
\label{bc}
\beta_2\de_2\vp-\beta_1\de_2\theta&=&0,\no
i\de_2\vp-\beta_1\de_1 Y&=&0,\no
i\de_2 Y+\beta_1\rho^2\de_1 \vp+\beta_2 r^2
\de_1\theta&=&0. 
\ea
In these equations the first equation indicates that 
the following condition should be satisfied on the world-volume of the 
D2-brane:
\ba
\label{constraint}
\beta_2\vp=\beta_1\theta+\mbox{constant}.
\ea
Then by comparing the result (\ref{bc}) 
to the general formula of the boundary condition (\ref{gbc}) with an 
additional constraint (\ref{constraint}), we can see the following flux 
on the world-volume of D2-branes 
\ba 
\label{flux}
\beta_1 F_{\vp Y}+\beta_2 F_{\theta Y}=1.
\ea
Then this flux is properly quantized on the world-volume of the D2-brane  
\ba
\frac{1}{4\pi^2\al}\int~F=\frac{1}{4\pi^2\al}\int~d\xi_1 d\xi_2~F_{\xi_1\xi_2}
=N,
\ea
where $\xi_1,\xi_2$ parameterize the world-volume of the D2-brane:
$\{(\vp,\theta,Y)~|~\beta_2\vp
=\beta_1\theta+\mbox{constant},~0\leq\vp<2\pi
k_1,~0\leq\theta<2\pi k_2\}$\footnote{This periodicity of $\vp$ and
$\theta$ is effective one which is only available for the world-volume
of a D2-brane.}. From this we can see that this system
represents a bound state of $N$ D0-branes and one D2-brane.

Moreover we can see the stabilization mechanism of the D2-brane 
by the analysis of the Dirac-Born-Infeld theory in the same way as 
(\ref{mass}). The total mass turns to be equal to that of $N$
D0-branes. Namely, 
the D2-brane part again becomes tensionless by the total effect of the NSNS
B-field and the magnetic flux $F$ (\ref{flux}). The analysis from the
world-volume theory of D0-branes is the same as before. We can see the
structure of the fuzzy torus with the noncommutativity $\theta=\frac{1}{N}$. 

\section{Conclusions}
\setcounter{equation}{0}
\hspace{5mm}

In this paper we have investigated the D-brane spectrum in the two parameter 
NSNS Melvin background from the viewpoint of both D-brane world-volume
theory and string world-sheet theory. 
In particular we have found that for non-zero rational values
of the B-field parameter 
$\beta\al/R=k/N$, neither pure D0-branes nor D2-branes can
exist except at the origin $\rho=0$. Instead we can put D2-D0 bound states for
any $\rho\neq0$ (see Fig.1). Interestingly, this object, which wraps 
a topologically trivial torus in the Melvin background, is indeed stabilized 
by the presence of $B$ and $F$ flux. The existence of it is also completely
consistent with the analysis of boundary states \cite{TU2} and indeed the 
bound state just corresponds to a system of $N$ fractional D0-branes in the
free field representation.
Furthermore, we have argued that 
this torus brane can be constructed from $N$ D0-branes 
by using the proposed non-abelian DBI-action of D0-branes.
This result is consistent with the fact that 
the effective world-volume theory on the D2-D0 bound state is identified with 
noncommutative torus (fuzzy torus) ${\ca{A}}_{\theta}\ \ (\theta=k/N)$.

In the same way we can also determine in the original 
Melvin background what kind of objects correspond to 
the other types of D-branes in the free field representation 
$(Y',\rho,\vp'')$ (see Table.1).
Furthermore we have shown that the similar kind of D2-D0 bound states
exists for the
higher dimensional Melvin background \cite{TU1,RuTs}. For the special 
values of the parameters we obtain BPS D2-D0 bound states.

\bigskip
\begin{center}
\noindent{\large \bf Acknowledgments}
\end{center}

We are grateful to Y. Michishita, S. Minwalla and P. Yi for showing us their 
interesting results and useful discussions. We also thank 
Y. Hikida, S. Kawamoto and T. Matsuo for discussions.
T.T. is supported by JSPS Research Fellowships for Young Scientists.


\begin{thebibliography}{99}
\baselineskip=13pt

\bibitem{Me}
M.A. Melvin, 
``Pure Magnetic and Electric Geons,''
Phys. Lett. B {\bf 8} (1964) 65.

\bibitem{CG}
M.~S.~Costa and M.~Gutperle,
``The Kaluza-Klein Melvin solution in M-theory,''
JHEP {\bf 0103} (2001) 027, hep-th/0012072.

\bibitem{Sa}
P.~M.~Saffin,
``Gravitating fluxbranes,''
Phys.\ Rev.\ D {\bf 64} (2001) 024014, gr-qc/0104014.

\bibitem{GS}
M.~Gutperle and A.~Strominger,
``Fluxbranes in string theory,''
JHEP {\bf 0106} (2001) 035, hep-th/0104136.

\bibitem{RT}
J.~G.~Russo and A.~A.~Tseytlin,
``Magnetic backgrounds and tachyonic instabilities in closed 
superstring  theory and M-theory,''
Nucl.\ Phys.\ B {\bf 611} (2001) 93, hep-th/0104238.

\bibitem{CoHeCo}
M. S. Costa, C. A.R. Herdeiro and L. Cornalba,
``Flux-branes and the Dielectric Effect in String Theory,''
hep-th/0105023.

\bibitem{Em}
R.~Emparan,
``Tubular branes in fluxbranes,''
Nucl.\ Phys.\ B {\bf 610} (2001) 169, hep-th/0105062.


\bibitem{Ur}
A.~M.~Uranga,
``Wrapped fluxbranes,'' hep-th/0108196.

\bibitem{MoTr}
J.~F.~Morales and M.~Trigiante,
``Walls from fluxes: An analytic RG-flow,'' hep-th/0112059.

\bibitem{BM3}
A.~A.~Tseytlin,
``Melvin solution in string theory,''
Phys.\ Lett.\ B {\bf 346} (1995) 55, hep-th/9411198.

\bibitem{BM}
J.~G.~Russo and A.~A.~Tseytlin,
``Exactly solvable string models of curved space-time backgrounds,''
Nucl.\ Phys.\ B {\bf 449} (1995) 91, hep-th/9502038.

\bibitem{SM}
J.~G.~Russo and A.~A.~Tseytlin,
``Magnetic flux tube models in superstring theory,''
Nucl.\ Phys.\ B {\bf 461} (1996) 131, hep-th/9508068.

\bibitem{TU1}
T.~Takayanagi and T.~Uesugi,
``Orbifolds as Melvin geometry,''
JHEP {\bf 0112} (2001) 004, hep-th/0110099.

\bibitem{RuTs}
J.G. Russo, A.A. Tseytlin,
``Supersymmetric fluxbrane intersections and closed string tachyons,''
hep-th/0110107.  

\bibitem{Su2}
T.~Suyama,
``Properties of String Theory on Kaluza-Klein Melvin Background,''
hep-th/0110077.

\bibitem{MiYi}
Y.~Michishita and P.~Yi,
``D-brane probe and closed string tachyons,'' hep-th/0111199.

\bibitem{Minwalla}
J.~R.~David, M.~Gutperle, M.~Headrick and S.~Minwalla,
``Closed string tachyon condensation on twisted circles,'' hep-th/0111212.

\bibitem{Su3}
T.~Suyama,
``Charged Tachyons and Gauge Symmetry Breaking,'' hep-th/0112101.

\bibitem{TU2}
T.~Takayanagi and T.~Uesugi,
``D-branes in Melvin background,''
JHEP {\bf 0111} (2001) 036, hep-th/0110200.

\bibitem{Ta}
W.~Taylor,
``D2-branes in B fields,''
JHEP {\bf 0007} (2000) 039, hep-th/0004141.


\bibitem{douglas}
C. Bachas, M. Douglas and C. Schweigert, 
``Flux stabilization of D-branes," JHEP {\bf 0005:048} (2000), hep-th/0003037.

\bibitem{paw}
J. Pawelczyk, ``SU(2) WZW D-branes and 
Their Noncommutative Geometry from DBI Action," hep-th/0003057.

\bibitem{ARS}
A.Y. Alekseev, A. Recknagel and V. Schomerus, ``Branes Dynamics in 
Background Fluxes and Non-Commutative Geometry," JHEP {\bf 0005:010} (2000), 
hep-th/0003187.

\bibitem{Ha}
K.~Hashimoto and K.~Krasnov,
``D-brane solutions in non-commutative gauge theory on fuzzy sphere,''
Phys.\ Rev.\ D {\bf 64} (2001) 046007, hep-th/0101145.

\bibitem{HNS}
Y.~Hikida, M.~Nozaki and Y.~Sugawara,
``Formation of spherical D2-brane from multiple D0-branes,''
Nucl.\ Phys.\ B {\bf 617} (2001) 117, hep-th/0101211.

\bibitem{Hyakutake}
Y.~Hyakutake,
``Expanded Strings in the Background of NS5-branes via a M2-brane, a 
D2-brane and D0-branes,'' hep-th/0112073.

\bibitem{Sen}
A.~Sen,
``Tachyon condensation on the brane antibrane system,''
JHEP {\bf 9808} (1998) 012, hep-th/9805170.

\bibitem{DoMo}
M.~R.~Douglas and G.~W.~Moore,
``D-branes, Quivers, and ALE Instantons,'' hep-th/9603167.

\bibitem{DiDoGo}
D.~E.~Diaconescu, M.~R.~Douglas and J.~Gomis,
``Fractional branes and wrapped branes,''
JHEP {\bf 9802} (1998) 013, hep-th/9712230.

\bibitem{CDS}
A.~Connes, M.~R.~Douglas and A.~Schwarz,
``Noncommutative geometry and matrix theory: Compactification on tori,''
JHEP {\bf 9802} (1998) 003, hep-th/9711162.

\bibitem{SW}
N.~Seiberg and E.~Witten,
``String theory and noncommutative geometry,''
JHEP {\bf 9909} (1999) 032, hep-th/9908142.


\bibitem{Ts0}
A.~A.~Tseytlin,
``On non-abelian generalization of the Born-Infeld action in string theory,''
Nucl.\ Phys.\ B {\bf 501} (1997) 41, hep-th/9701125.

\bibitem{My1}
R.~C.~Myers,
``Dielectric-branes,''
JHEP {\bf 9912} (1999) 022, hep-th/9910053.

\bibitem{My2}
N.~R.~Constable, R.~C.~Myers and O.~Tafjord,
``The noncommutative bion core,''
Phys.\ Rev.\ D {\bf 61} (2000) 106009, hep-th/9911136;
N.~R.~Constable, R.~C.~Myers and O.~Tafjord,
``Non-Abelian brane intersections,''
JHEP {\bf 0106} (2001) 023, hep-th/0102080;
N.~R.~Constable, R.~C.~Myers and O.~Tafjord,
``Fuzzy funnels: Non-abelian brane intersections,'' hep-th/0105035;
R.~C.~Myers,
``Nonabelian D-branes and noncommutative geometry,''
Int.\ J.\ Mod.\ Phys.\ A {\bf 16}, 956 (2001), hep-th/0106178.

\bibitem{DuMo}
E.~Dudas and J.~Mourad,
``D-branes in string theory Melvin backgrounds,'' hep-th/0110186.

\end{thebibliography}
\end{document}